\documentclass[twocolumn,pra,aps]{revtex4}

\begin{document}
\title{Nonlocality in unambiguous pure-state identification without classical knowledge} 
\author{A.~Hayashi, Y.~Ishida, T.~Hashimoto, and M.~Horibe}
\affiliation{Department of Applied Physics\\
           University of Fukui, Fukui 910-8507, Japan}

\begin{abstract}
For two bipartite pure states, we consider the problem of unambiguous identification 
without classical knowledge on the states. The optimal success probability by means 
of local operations and classical communication is shown to be less than the maximum 
attainable by the global measuring scheme.    
\end{abstract}

\maketitle

\newcommand{\ket}[1]{|\,#1\,\rangle}
\newcommand{\bra}[1]{\langle\,#1\,|}
\newcommand{\braket}[2]{\langle\,#1\,|\,#2\,\rangle}
\newcommand{\bold}[1]{\mbox{\boldmath $#1$}}
\newcommand{\sbold}[1]{\mbox{\boldmath ${\scriptstyle #1}$}}
\newcommand{\tr}[1]{{\rm tr}\!\left[#1\right]}
\newcommand{\trm}{{\rm tr}}
\newcommand{\BC}{{\bold{C}}}
\newcommand{\CS}{{\cal S}}
\newcommand{\CA}{{\cal A}}
\newcommand{\CM}{{\cal M}}

\section{Introduction}
Distinguishing different quantum states by measurement is one of the most fundamental 
and important problems in quantum information theory \cite{Helstrom76,Holevo82}. 
Two features of quantum mechanics make this problem extremely nontrivial. 
One is the statistical nature of quantum measurement. To obtain complete information 
on a given state, we need unlimited number of copies of the state, since an unknown quantum 
state cannot be cloned \cite{Wootters82}. The other feature is nonlocality in quantum 
mechanics, which manifests itself typically when the entangled states are involved. 
Even for separable multipartite states, the global measurement on the whole system sometimes 
performs better than the scheme based on local operations and classical 
communication (LOCC) \cite{Peres91,Ban97,Sasaki98,Eldar01,Bennett99,Koashi07}.   

The problem we consider here is the unambiguous identification of two bipartite pure 
states. Alice and Bob are given a bipartite pure state, which is guaranteed 
to be one of the two pure reference states. 
Contrary to the standard unambiguous (conclusive) discrimination problem 
\cite{Ivanovic87,Dieks88,Peres88}, Alice and Bob are not given any classical information 
on the two reference states. Instead, a certain number ($=N$) of copies of the reference states 
are available to Alice and Bob. Furthermore, each of the two reference states are randomly 
chosen, therefore, generally entangled. The task of Alice and Bob is unambiguously 
identify the given input state with one of the two reference states by means of an LOCC 
scheme. The question is whether Alice and Bob can achieve the maximum mean success 
probability attainable by the global measurement scheme. 
For the global identification problem, see \cite{Hayashi05,Bergou05,Hayashi06}.

When the number $N$ of copies of the reference states is infinity, 
one can in principle obtain complete classical knowledge on the states. 
Note that this can be done within LOCC schemes. In this limit, 
the problem reduces to the standard discrimination problem of two 
pure bipartite states. For those discrimination problems with classical knowledge on the 
states assumed, several results have been known. 
First, Walgate {\it et al}. \cite{Walgate00} showed that any two mutually orthogonal pure 
states can be perfectly distinguished by LOCC, regardless of entanglement of the states.  
Then, it was shown that any two generally nonorthogonal pure states 
can be optimally discriminated by LOCC. This was shown for the two types of  
discrimination problems: the inconclusive discrimination problem \cite{Virmani01} 
where error is allowed and the unambiguous (conclusive) discrimination 
problem \cite{Ji05,Chen01,Chen02} where no error is allowed but an inconclusive 
guess can be made.  
 
We recently studied the inconclusive identification problem of two bipartite pure states for 
the case of $N=1$ and demonstrated that the LOCC based scheme can achieve the maximum 
success probability attainable by the global measurement scheme \cite{Ishida08}.  

In this paper, we consider the unambiguous (conclusive) identification of two bipartite pure 
states in the case of $N=1$, where no error is allowed but an inconclusive guess can be made. 
We will show that the globally attainable optimal success probability cannot be achieved by 
any LOCC based scheme.    

\section{Unambiguous identification and symmetries of POVM}
In this section, we precisely formulate the unambiguous identification problem of two pure 
states \cite{Hayashi06}, and determine the optimal success probability attainable by 
the global measurement scheme. In doing so, we explain two important symmetries 
of the measurement scheme of this problem, which will also play the crucial role in 
determining the optimal probability by LOCC scheme in the next section.  

\subsection{Problem}
We have three systems 0, 1, and 2, each on a $d$-dimensional complex vector space $\BC^d$.  
The input state $\rho=\ket{\phi}\bra{\phi}$ is prepared in system 0, 
whereas two pure reference states $\rho_1=\ket{\phi_1}\bra{\phi_1}$ and $\rho_2=\ket{\phi_2}\bra{\phi_2}$ 
is prepared in system 1 and 2, respectively. It is promised that the input state is 
equal to one of the two reference states with equal probabilities.  We do not have 
any classical knowledge on the reference states. 
Instead, the two reference states are independently chosen from the state space $\BC^{d}$ in  
a unitary invariant way. More precisely, the distribution is assumed uniform 
on a $2d-1$ dimensional unit hypersphere of $2d$ real variables 
$\{{\rm Re} c_i, {\rm Im} c_i\}_{i=0}^{d-1}$, where $c_i$ is expansion coefficients of 
the state in terms of an orthonormal basis on $\BC^{d}$ \cite{Hayashi04}.
Our task is to unambiguously identify the input state with one of the two reference states. 
Our measurement can produce three outcomes $\mu=1,2,0$. If the outcome is $\mu(=1,2)$, we are certain 
that the input state $\rho$ is $\rho_\mu$, and outcome 0 means that we are not certain 
about the identity of the input, which is called the inconclusive result. 

Let us introduce a positive operator-valued measure (POVM), $\{E_\mu \}_{\mu=0,1,2}$, 
corresponding to the three measurement outcomes defined above.   
The mean success probability of identification is then given by
\begin{eqnarray}
  p &=& \frac{1}{2} \Big< \tr{E_1\rho_1(0)\rho_1(1)\rho_2(2)}
                  \nonumber \\
    & & \hspace{8ex} +\tr{E_2\rho_2(0)\rho_1(1)\rho_2(2)} \Big>, 
                   \label{p1}
\end{eqnarray}
where the symbol $<\cdots>$ represents the average over the two reference states, and 
we specify the system which an operator acts on by the system number (0, 1, 2) in the parentheses; 
$\rho(0)$ is the operator on system 0, for example.  The condition that we are not allowed to 
make a mistake imposes the following no-error conditions on $E_1$ and $E_2$: 
\begin{eqnarray}
   \left\{ 
   \begin{array}{c}
      \tr{E_1\rho_2(0)\rho_1(1)\rho_2(2)} = 0, \\
      \tr{E_2\rho_1(0)\rho_1(1)\rho_2(2)} = 0, \\ 
   \end{array}
   \right. 
   \ \ \mbox{for any $\rho_1$ and $\rho_2$}.
                          \label{no-error1}
\end{eqnarray}
In what follows, we will optimize the mean success probability of Eq.(\ref{p1}) under the 
no-error conditions. 

The average over the reference states can be performed by using the formula \cite{Hayashi04}:
\begin{eqnarray}
  < \rho^{\otimes n} > = \frac{\CS_n}{d_n},
\end{eqnarray}
where $\CS_n$ is the projector onto the totally symmetric subspace of $(\BC^d)^{\otimes n}$ 
and $d_n$ is its dimension given by $d_n={}_{n+d-1}C_{d-1}$.

The mean success probability (\ref{p1}) then takes the form:
\begin{eqnarray}
  p(d) = \frac{1}{2d_2d_1}\left( \tr{E_1\CS(01)} + \tr{E_2\CS(02)} \right), 
           \label{pS}
\end{eqnarray}
where $\CS(01)$ and $\CS(02)$ are the projector onto the totally symmetric subspace of 
space $0 \otimes 1$ and $0 \otimes 2$, respectively.  
Averaging the no-error conditions of Eq.(\ref{no-error1}), we obtain
\begin{eqnarray}
  \tr{E_1\CS(02)} = 0,\ \ \tr{E_2\CS(01)}=0.    \label{no-error2}
\end{eqnarray}
Since $E_1$ and $\CS(02)$ are both positive operators, the above conditions imply 
the supports of them are orthogonal to each other. The same is true for 
$E_2$ and $\CS(01)$. The no-error conditions are thus equivalent to 
\begin{eqnarray}
  && E_1 \CS(02) = \CS(02) E_1 = 0, \label{no-error31} \\
  && E_2 \CS(01) = \CS(01) E_2 = 0. \label{no-error32}
\end{eqnarray}

\subsection{Symmetries of POVM}
The set of POVM's satisfying the no-error conditions is convex; 
if two POVM's $E_\mu$ and $E'_\mu$ respect the no-error conditions, 
so does their convex linear combination $qE_\mu+(1-q)E'_\mu$ for any 
$ 0 \le q \le 1$.
The resulting success probability is also a convex combination: 
$p(qE+(1-q)E') = qp(E) + (1-q)p(E')$ with an obviously abbreviated notation. 
It is this convexity of POVM that we exploit in order to impose some symmetries 
on the optimal POVM without loss of generality. 

First we consider the exchange symmetry between systems 1 and 2.
For an optimal POVM $F_\mu$, we define another POVM by 
\begin{eqnarray}
  && F'_1 = T(12) F_2 T(12),   \nonumber \\
  && F'_2 = T(12) F_1 T(12),   \nonumber \\
  && F'_0 = T(12) F_0 T(12),   \nonumber 
\end{eqnarray}
where $T(12)$ is the exchange operator between systems 1 and 2.
The POVM $F'_\mu$ is clearly legitimate and optimal. Furthermore,
a new POVM $ E_\mu = \frac{1}{2} \left( F_\mu + F'_\mu \right) $, 
which is a convex linear combination of $F_\mu$ and $F'_\mu$, 
is also optimal and satisfies the exchange symmetry between systems 1 and 2:
\begin{eqnarray}
   && E_1=T(12)E_2T(12),   \nonumber \\
   && E_2=T(12)E_1T(12),   \nonumber \\
   && E_0=T(12)E_0T(12).   \label{exchange_symmetry}
\end{eqnarray}

The second important symmetry is the unitary symmetry of the 
distribution of the reference states. If a POVM $F_\mu$ is optimal, 
another POVM defined by 
\begin{eqnarray}
   U^{\otimes 3}F_\mu (U^{\otimes 3})^{-1}, 
\end{eqnarray} 
is also legitimate and optimal for arbitrary unitary operator $U$.
We now construct a POVM by 
\begin{eqnarray}
    E_\mu = \int dU U^{\otimes (3)} F_\mu (U^{\otimes (3)})^{-1}, 
\end{eqnarray}
where $dU$ is the normalized positive invariant measure of the group $U(d)$.
The POVM $E_\mu$ is clearly a legitimate and optimal POVM.  
We can show that $E_\mu$ commutes with $U^{\otimes 3}$ for any unitary $U$:
\begin{eqnarray}
   U^{\otimes 3} E_\mu &=&  \int dU' (UU')^{\otimes (3)} F_\mu (UU'^{\otimes (3)})^{-1}U^{\otimes 3}
                         \nonumber \\
     &=& \int dU' U'^{\otimes (3)} F_\mu (U'^{\otimes (3)})^{-1} U^{\otimes 3}
                         \nonumber \\
     &=& E_\mu U^{\otimes 3},
\end{eqnarray}
which means that $E_\mu$ is a scalar with respect to the group $U(d)$. 
Thus we can assume that the optimal POVM satisfies the exchange symmetry of 
Eq.(\ref{exchange_symmetry}) and is scalar with respect to the group $U(d)$.

By the exchange symmetry, the mean success probability to be optimized 
takes the form:
\begin{eqnarray}
 p = \frac{1}{d_2d_1}\tr{E_1\CS(01)}, 
                                       \label{pS_symmetric}
\end{eqnarray}
where $E_1$ is a unitary scalar and subject to the conditions:
\begin{eqnarray}
  E_1 \ge 0,\ 1 \ge E_1+T(12)E_1T(12), 
                 \label{positivity_symmetric}
\end{eqnarray}
in addition to the no-error conditions given by Eq.(\ref{no-error31}).

\subsection{Optimal identification probability}
Let us decompose the total space into three subspaces according to the symmetry with 
respect to system permutations. 
\begin{eqnarray}
  \BC^d \otimes \BC^d \otimes \BC^d=V_\CS \oplus V_\CA \oplus V_\CM.
\end{eqnarray}
Here $V_\CS$ is the totally symmetric subspace of $\dim V_\CS \equiv d_3=d(d+1)(d+2)/6$ 
and $V_\CA$ is the totally antisymmetric subspace of $\dim V_\CM =d(d-1)(d-2)/6$. 
The remaining subspace $V_\CM$ is the mixed symmetric subspace of $\dim V_\CM =2d(d^2-1)/3$.  
The subspace $V_\CM$ contains the 2-dimensional irreducible representation of the symmetric group 
of order 3 with multiplicity $\dim V_\CM /2$. 
We denote projectors onto 
$V_\CS$, $V_\CA$, and $V_\CM$ by $\CS_3$, $\CA_3$ and $\CM_3$, respectively. 

Under the unitary transformation $U^{\otimes 3}$, the three subspaces $V_\CS$, $V_\CA$, and $V_\CM$ 
are clearly invariant since $U^{\otimes 3}$ commutes with system permutations.  
Furthermore, it is known that $U^{\otimes 3}$ acts on $V_\CS$ and $V_\CA$ irreducibly, 
and the mixed symmetric space $V_\CM$ contains two $(\dim V_\CM /2)$-dimensional irreducible 
representations of the group $U(d)$ \cite{Hamermesh62}. 
Now, suppose that a positive operator $E$ is a unitary scalar; $E$ commutes with 
$U^{\otimes 3}$. If $\tr{E\CS(02)}=0$, $E$ should be  a linear combination of 
two projection operators $\CA_3$ and $\CM_3 \CA(02)$, which is a consequence of Schur's lemma. 
Similarly, if $\tr{E\CA(02)}=0$, $E$ is a linear combination of $\CS_3$ and $\CM_3 \CS(02)$. 
These facts will be used also in the next section.

With these considerations, we can determine the operator form of POVM element $E_1$.  
The operator $E_1$ is given by a linear combination of $\CA_3$ and $\CM_3 \CA(02)$ since 
$E_1$ is a unitary scalar and satisfies the no-error condition $\tr{E_1\CS(02)}=0$. 
But $\CA_3$ does not contribute to the mean success probability of Eq.(\ref{pS_symmetric}). 
Thus, without loss of generality, we can write 
\begin{eqnarray}
  E_1 = \alpha \CM_3 \CA(02),
\end{eqnarray}
where $\alpha$ is a positive coefficient. The range of $\alpha$ is restricted by the positivity 
of $E_0$.

Here, it is convenient to introduce two operators $D$ and $A$ as  
\begin{eqnarray}
    D &\equiv& \frac{1}{2} \left( T(01) - T(02) \right),  \label{D} \\
    A &\equiv& \frac{1}{2} \left(  T(01) + T(02) \right). \label{A}
\end{eqnarray}
Calculating $D^2$, we find 
\begin{eqnarray}
   D^2 &=& \frac{1}{4} \left( 2-T(01)T(02)-T(02)T(01) \right)  \nonumber \\
       &=& \frac{3}{4} \left(1 - \CS_3 -\CA_3 \right)          \nonumber \\
       &=& \frac{3}{4} \CM_3,
\end{eqnarray}
which implies that eigenvalues of $D$ are $\pm \sqrt{3}/2$ in $V_\CM$ and 0 otherwise. 
It is also easy to show that 
\begin{eqnarray} 
  && A^2 + D^2= 1, \label{asquare}  \\
  && DA+AD=0. \label{anti} 
\end{eqnarray}
From Eq.(\ref{asquare}), it is clear that $A$ has eigenvalues $\pm 1/2$ in $V_\CM$. 
In $V_\CM$, each eigenvalue of $D$ and $A$ has the same multiplicity $\dim V_\CM/2$, 
since the anticommutation relation of Eq.(\ref{anti}) shows that each of the operators 
$D$ and $A$ changes the sign of eigenvalue of the other. 

We can now determine the range of $\alpha$ by the positivity of $E_0$:
\begin{eqnarray}
    E_1 + T(12)E_1T(12) &=& \alpha\CM_3(\CA(02)+\CA(01))   \nonumber \\
                        &=& \alpha\CM_3\left( 1+A \right) \le 1.
\end{eqnarray}
This requires $\alpha \le 2/3 $, since $A$ has eigenvalues $\pm 1/2$ in $V_\CM$. 
Clearly, the mean success probability attains its maximum when $\alpha$ takes the 
largest possible value $2/3$. The optimal POVM is thus given by 
\begin{eqnarray}
  E_1 &=& \frac{2}{3}\CM_3\CA(02),   \nonumber \\
  E_2 &=& \frac{2}{3}\CM_3\CA(01),  \nonumber \\
  E_0 &=& \frac{1}{3}\CM_3(1+2A)+\CS_3+\CA_3. 
\end{eqnarray}
In order to obtain the optimal probability, we need trace $\tr{\CM_3\CA(02)\CS(01)}$, 
which is calculated as follows: 
\begin{eqnarray}
 &&  \tr{\CM_3\CA(02)\CS(01)}     \nonumber \\
 &=& \frac{1}{4}\tr{\CM_3 (1+T(01)-T(02)-T(02)T(01))}  \nonumber \\
 &=& \frac{1}{2}\tr{\CM_3 D^2} = \frac{3}{8}\dim V_\CM, 
\end{eqnarray}
where we used $\tr{\CM_3T(01)}=\tr{\CM_3T(02)}=0$.
Using the explicit expressions for the dimensions, we finally obtain 
the optimal mean success probability of unambiguous identification:
\begin{eqnarray}
  p_{\max}= \frac{d-1}{3d}. \label{global_p}
\end{eqnarray}

\section{Local unambiguous identification}
Let us now assume that each of the three systems consists of two subsystems shared by 
Alice and Bob and its state space is represented by a tensor product 
$\BC^d=\BC^{d_a} \otimes \BC^{d_b}$.  
The two bipartite reference states $\ket{\phi_1}$ and $\ket{\phi_2}$ are independently chosen 
according to the unitary invariant distribution on $\BC^d$ as in the preceding section. 
Therefore, they are generally entangled. 
The task of Alice and Bob is to unambiguously identify a given input state 
by means of local operations and classical communication (LOCC) with one of the two reference states. 

\subsection{Separable POVM and symmetries}
Any POVM $E_\mu^{{\rm L}}$ which satisfies the LOCC conditions has a separable 
form:  
\begin{eqnarray}
    E_\mu^{{\rm L}} = \sum_i E_{\mu i}^{(a)} \otimes E_{\mu i}^{(b)},\ \ \ (\mu=0,1,2), 
\end{eqnarray}
where
\begin{eqnarray}
     E_{\mu i}^{(a)} \ge 0, \ \ E_{\mu i}^{(b)} \ge 0,\ \ \sum_\mu E_\mu^{{\rm L}} = 1.
\end{eqnarray}
Here and hereafter the superscript $(a)$ or $(b)$ of an operator indicates which space of 
Alice or Bob the operator acts on, 
and the symbol of tensor product will sometimes be omitted.  
It is known that there exist separable POVM's which do not satisfy the LOCC conditions \cite{Bennett99}. 
We will first optimize the success probability within the separable class of POVM, and 
then show that the obtained optimal separable POVM can be implemented by an LOCC protocol.
 
Note that a convex linear combination of separable POVM's is again separable, and the no-error 
conditions $\tr{E_1^{{\rm L}}\CS(02)}=\tr{E_2^{{\rm L}}\CS(01)}=0$ are also preserved. 
This enables us to impose two symmetries on the optimal separable POVM as in the preceding 
section. We begin with the exchange symmetry for system 1 and 2. Since 
$T(12)=T^{(a)}(12) \otimes T^{(b)}(12)$, it is clear that $T(12)E_\mu^{{\rm L}} T(12)$ is separable if 
$E_\mu^{{\rm L}}$ is separable:
\begin{eqnarray*}
  && T(12)E_\mu^{{\rm L}} T(12)   \nonumber \\ 
  &=& \sum_i T^{(a)}(12)E_{\mu i}^{(a)}T^{(a)}(12) \otimes 
                             T^{(b)}(12)E_{\mu i}^{(b)}T^{(b)}(12).
\end{eqnarray*}
Therefore, we can impose on separable POVM the same exchange symmetry to the one given in 
Eq.(\ref{exchange_symmetry}):
\begin{eqnarray}
  E_1^{{\rm L}} &=& T(12) E_2^{{\rm L}} T(12), \nonumber \\
  E_2^{{\rm L}} &=& T(12) E_1^{{\rm L}} T(12), \nonumber \\
  E_0^{{\rm L}} &=& T(12) E_0^{{\rm L}} T(12). 
\end{eqnarray}

For the unitary symmetry, we notice that $U^{\otimes 3} F_\mu^{{\rm L}} (U^{\otimes 3})^{-1}$ is not 
generally separable for a separable POVM 
$F_\mu^{{\rm L}} = \sum_i F_{\mu i}^{(a)} \otimes F_{\mu i}^{(b)}$. 
However, this is true if $U$ is a tensor product of two unitaries as $U=u^{(a)} \otimes v^{(b)}$:
\begin{eqnarray*}
   & & U^{\otimes 3} F_\mu^{{\rm L}} (U^{\otimes 3})^{-1} \nonumber \\
   &=& \sum_i  u^{(a)\otimes 3} F_{\mu i}^{(a)} (u^{(a)\otimes 3})^{-1} \otimes
            v^{(b)\otimes 3} F_{\mu i}^{(b)} (v^{(b)\otimes 3})^{-1}.
\end{eqnarray*}
For the class of this separable $U$, we can repeat the argument given in the preceding section.
Assume a separable POVM $F_\mu^{{\rm L}} $ is optimal.  
Integrating over $u^{(a)}$ and $v^{(b)}$ with the invariant measure, we obtain 
\begin{eqnarray}
   E_\mu^{{\rm L}} &\equiv& \int du^{(a)}dv^{(b)}\left(u^{(a) \otimes 3} v^{(b) \otimes 3} \right)
                       F_\mu^{{\rm L}}    \left(u^{(a) \otimes 3} v^{(b) \otimes 3} \right)^{-1}  
                                    \nonumber \\
         &=& \sum_i E_{\mu i}^{(a)}\otimes E_{\mu i}^{(b)}, 
\end{eqnarray}
where $E_{\mu i}^{(a)}$ and $E_{\mu i}^{(b)}$ are given by 
\begin{eqnarray}
   E_{\mu i}^{(a)} &=& \int du^{(a)} u^{(a)\otimes 3} F_{\mu i}^{(a)} (u^{(a)\otimes 3})^{-1}, 
                        \nonumber \\
   E_{\mu i}^{(b)} &=& \int dv^{(b)} v^{(b)\otimes 3} F_{\mu i}^{(b)} (v^{(b)\otimes 3})^{-1}. 
\end{eqnarray} 
The POVM $ E_\mu^{{\rm L}} $ obtained this way is again separable and optimal. 
Furthermore, it is easy to see that $E_{\mu i}^{(a)}$ and $E_{\mu i}^{(b)}$ are both 
unitary scalar: for any unitaries $u^{(a)}$ and $v^{(b)}$, we have 
\begin{eqnarray}
    [E_{\mu i}^{(a)}, u^{(a)}] = 0,\ \ \ [E_{\mu i}^{(b)}, v^{(b)}]=0.
\end{eqnarray}

Let us closely examine the no-error conditions for separable POVM. 
The global projector $\CS(02)$ is decomposed by local symmetry projectors as 
follows:                        
\begin{eqnarray*}
  \CS(02) = \CS^{(a)}(02) \otimes \CS^{(b)}(02)+\CA^{(a)}(02) \otimes \CA^{(b)}(02).
\end{eqnarray*}
The no-error condition $\tr{E_1^{{\rm L}}\CS(02)}=0$ is then expressed as  
\begin{eqnarray}
  && \sum_i \Big( \tr{E_{1 i}^{(a)}\CS^{(a)}(02)}\tr{E_{1 i}^{(b)}\CS^{(b)}(02)} 
                           \nonumber \\
  && \hspace{2ex} +\tr{E_{1 i}^{(a)}\CA^{(a)}(02)}\tr{E_{1 i}^{(b)}\CA^{(b)}(02)} 
            \Big) = 0.
\end{eqnarray}                     
In this equation, all terms are non-negative, implying each term should vanish. 
Therefore, for each $i$, we have two possibilities: one is 
\begin{eqnarray}
   \tr{E_{1 i}^{(a)}\CS^{(a)}(02)}=0,\ \mbox{and}\ \tr{E_{1 i}^{(b)}\CA^{(b)}(02)} =0,
                       \label{one_possibility}
\end{eqnarray}
and the other is 
\begin{eqnarray}
   \tr{E_{1 i}^{(a)}\CA^{(a)}(02)}=0,\ \mbox{and}\ \tr{E_{1 i}^{(b)}\CS^{(b)}(02)} =0.
                       \label{other_possibility}
\end{eqnarray}
Note that the other combinations like 
\begin{eqnarray*}
   \tr{E_{1 i}^{(a)}\CS^{(a)}(02)}=0,\ \mbox{and}\ \tr{E_{1 i}^{(a)}\CA^{(a)}(02)} =0,
\end{eqnarray*}
do not occur as this would imply $E_{1 i}^{(a)}$ or $E_{1 i}^{(b)}$ is identically zero. 
From the no-error condition for $E_2^{{\rm L}}$, we obtain the similar conditions 
for its components $E_{2 i}^{(a)}$ and $E_{2 i}^{(b)}$.

\subsection{Possible operator form of separable POVM}

As in the preceding section, we can show that a positive operator $E^{(p)}$ on space $V^{(p)}(p=a,b)$
which is a unitary scalar and satisfies $\tr{E^{(p)}\CS^{(p)}(02)}=0$ 
is a linear combination of $\CA_3^{(p)}$ and $\CM_3^{(p)}\CA^{(p)}(02)$. Similarly, 
if $E^{(p)}$ satisfies $\tr{E^{(p)}\CA^{(p)}(02)}=0$, then 
$E^{(p)}$ can be written as a linear combination of $\CS_3^{(p)}$ and $\CM_3^{(p)}\CS^{(p)}(02)$.
Now we can write the possible form of separable $E_1^{{\rm L}}$ which has the unitary symmetry and 
satisfies the no-error conditions: 
\begin{eqnarray}
   E_1^{{\rm L}} &=& 
       \alpha_1 \CS_3^{(a)} \otimes \CM_3^{(b)}\CA^{(b)}(02)
      +\alpha_2 \CA_3^{(a)} \otimes \CM_3^{(b)}\CS^{(b)}(02)   
                               \nonumber \\
   &+& \alpha_3 \CM_3^{(a)}\CS^{(a)}(02) \otimes \CA_3^{(b)}
      +\alpha_4 \CM_3^{(a)}\CA^{(a)}(02) \otimes \CS_3^{(b)}  
                               \nonumber \\
   &+& \beta_1 \CM_3^{(a)}\CS^{(a)}(02) \otimes \CM_3^{(b)}\CA^{(b)}(02)    
                               \nonumber \\              
   &+& \beta_2 \CM_3^{(a)}\CA^{(a)}(02) \otimes \CM_3^{(b)}\CS^{(b)}(02),  
                        \label{possible_E1}
\end{eqnarray}
where $\alpha_1,\alpha_2,\alpha_3,\alpha_4$ and $\beta_1,\beta_2$ are non-negative 
coefficients. The operators $\CS_3 \otimes \CA_3$ and $\CA_3 \otimes \CS_3$ are not 
included in $E_1^{{\rm L}}$, since the corresponding outcomes do not arise; the total state  
contains no totally antisymmetric component.  
$E_2^{{\rm L}}$ and $E_0^{{\rm L}}$ are given by 
$E_2^{{\rm L}} = T(12)E_1^{{\rm L}}T(12)$ and $E_0^{{\rm L}}=1-E_1^{{\rm L}}-E_2^{{\rm L}}$.

We have now found the possible operator form of separable POVM which respects the no-error conditions. 
The remaining requirement on the POVM is the positivity of $E_0^{{\rm L}}$, which restricts 
the range of the coefficients $\alpha$'s and $\beta$'s. The positivity of $E_0^{{\rm L}}$ 
is equivalent to $E_1^{{\rm L}}+E_2^{{\rm L}} \le 1$. We will separately check this inequality 
in each of all subspaces of the permutation symmetry. 

In the subspace $V_{\CS}^{(a)} \otimes V_{\CM}^{(b)}$, the relevant part of $E_1^{{\rm L}}+E_2^{{\rm L}}$ 
is written as 
\begin{eqnarray}
  && \alpha_1 \left( \CS_3^{(a)} \otimes \CM_3^{(b)}\CA^{(b)}(02) +
                      \CS_3^{(a)} \otimes \CM_3^{(b)}\CA^{(b)}(01)  \right)
                                 \nonumber \\
  &&=\alpha_1 \CS_3^{(a)} \otimes \CM_3^{(b)} (1-A^{(b)}) .
\end{eqnarray}
This part should be smaller than the projector onto the subspace, $\CS_3^{(a)} \otimes \CM_3^{(b)}$. 
Here, $A^{(b)}$ is defined to be $(T^{(b)}(01)+T^{(b)}(02))/2$ in the same way for the global operator $A$ 
introduced in the preceding section.  Eigenvalues of $A^{(b)}$ in $V_{\CM}^{(b)}$ are $1/2$ and $-1/2$. 
Therefore, we obtain $\alpha_1 \le 2/3$. Similarly, we obtain $\alpha_4 \le 2/3$ from the inequality 
in the subspace $V_{\CM}^{(a)} \otimes V_{\CS}^{(b)}$.

The inequality $E_1^{{\rm L}}+E_2^{{\rm L}} \le 1$ in $V_{\CA}^{(a)} \otimes V_{\CM}^{(b)}$ 
takes the form:  
\begin{eqnarray}
  & & \alpha_2 \left( \CA_3^{(a)} \otimes \CM_3^{(b)}\CS^{(b)}(02) +
                      \CA_3^{(a)} \otimes \CM_3^{(b)}\CS^{(b)}(01)  \right)
                                 \nonumber \\
  & &=\alpha_2 \CA_3^{(a)} \otimes \CM_3^{(b)} (1+A^{(b)}) \le \CA_3^{(a)} \otimes \CM_3^{(b)},  
\end{eqnarray} 
which requires that $\alpha_2 \le 2/3$.  In the same way, we obtain $\alpha_3 \le 2/3$  
from the inequality in $V_{\CM}^{(a)} \otimes V_{\CA}^{(b)}$. Thus all the four coefficients $\alpha$'s 
should be less or equal to 2/3. 

It is not straightforward to find allowed ranges of $\beta_1$ and $\beta_2$ from 
the inequality in $V_\CM^{(a)} \otimes V_\CM^{(b)}$.  
In the space $V_\CM^{(a)} \otimes V_\CM^{(b)}$, we define an operator $X$ to be 
\begin{eqnarray}
  X \equiv \beta_1 \left( \CS^{(a)}(02) \CA^{(b)}(02) 
                           +\CS^{(a)}(01) \CA^{(b)}(01) \right)\  &&
                                      \nonumber \\
   +\beta_2 \left( \CA^{(a)}(02) \CS^{(b)}(02) 
                            +\CA^{(a)}(01) \CS^{(b)}(01) \right), &&
\end{eqnarray}
which is the part of $E_1^{{\rm L}}+E_2^{{\rm L}}$ which contributes to $V_\CM^{(a)} \otimes V_\CM^{(b)}$. 
We need to find the greatest eigenvalue of $X$, since the inequality implies 
$X \le \CM_3^{(a)} \otimes \CM_3^{(b)}$. 
It should be understood that we are working in subspace $V_\CM^{(a)} \otimes V_\CM^{(b)}$, and 
the projectors $\CM_3^{(a)}$ and $\CM_3^{(b)}$ will be omitted. 
In terms of $A^{(p)}$ and $D^{(p)}$, the operator $X$ is expressed as 
\begin{eqnarray}
  X = \beta \left( 1 - A^{(a)} \otimes A^{(b)} - D^{(a)} \otimes D^{(b)} \right)\ &&
                      \nonumber \\
     +\delta \left( 1^{(a)} \otimes A^{(b)} - A^{(a)} \otimes 1^{(b)} \right), &&
\end{eqnarray}
where $\beta=\frac{1}{2}(\beta_1+\beta_2)$ and $\delta=\frac{1}{2}(\beta_1-\beta_2)$. 
In order to diagonalize $X$, it is convenient to introduce the basis in which $A^{(p)}(p=a,b)$ 
is diagonal:
\begin{eqnarray}
  A^{(p)}\ket{m+} = \frac{1}{2} \ket{m+},\ A^{(p)}\ket{m-} = -\frac{1}{2}\ket{m-},  &&  
                   \nonumber \\
    D^{(p)}\ket{m+} = \frac{\sqrt{3}}{2}\ket{m-},\ D^{(p)}\ket{m-} = \frac{\sqrt{3}}{2}\ket{m+},  &&  
\end{eqnarray}  
where $m=1,2,\ldots,\dim (V_\CM^{(p)})/2$. The bipartite state 
$\ket{m\pm} \otimes \ket{m'\pm}$ for a given set of $m$ and $m'$ will be 
written as $\ket{\!\pm\pm}$ for simplicity.   

In this basis, two eigenvalues of $X$ are easily found by inspection:
\begin{eqnarray*}
  && X ( \ket{\!++}+\ket{\!--} ) = 0,  \\
  && X ( \ket{\!++}-\ket{\!--} ) = \frac{3}{2}\beta ( \ket{\!++}-\ket{\!--} ).
\end{eqnarray*}
States $\ket{\!+-}$ and $\ket{\!-+}$ are transformed by $X$ as 
\begin{eqnarray}
  X \ket{\!+-} &=& \left(\frac{5}{4}\beta-\delta \right) \ket{\!+-} 
                         -\frac{3}{4}\beta\ket{\!-+} ,
                                           \nonumber \\ 
  X \ket{\!-+} &=& \left(\frac{5}{4}\beta+\delta \right) \ket{\!-+} 
                         -\frac{3}{4}\beta\ket{\!+-}.
\end{eqnarray}
The other two eigenvalues are determined by diagonalizing the 2 by 2 matrix corresponding to 
the above transformation and found to be  
\begin{eqnarray}
     \lambda_{\pm} = \frac{5}{4}\beta \pm \sqrt{ \frac{9}{16}\beta^2+\delta^2}.
\end{eqnarray}
Of the four eigenvalues, the greatest one is $\lambda_+$. 
The positivity of $E_0^{{\rm L}}$ thus requires that the positive coefficients $\beta_1$ and 
$\beta_2$ should satisfy the condition:
\begin{eqnarray}
  \frac{5}{4}\beta + \sqrt{ \frac{9}{16}\beta^2+\delta^2} \le 1, 
\end{eqnarray}
where $\beta=\frac{1}{2}(\beta_1+\beta_2)$ and $\delta=\frac{1}{2}(\beta_1-\beta_2)$.

\subsection{Maximum success probability by separable POVM}
Now that we have the possible form of separable POVM $E_\mu^{{\rm L}}$ and the conditions 
for the coefficients in it, we can optimize the mean success probability given by 
\begin{eqnarray}
  p^{{\rm L}} = \frac{1}{d_2d_1} \tr{E_1^{{\rm L}}\CS(01)}.
\end{eqnarray}
The trace $\tr{E_1^{{\rm L}}\CS(01)}$ can be calculated by decomposing it into traces in subsystems as  
\begin{eqnarray*}
   & & \tr{E_1^{{\rm L}}\CS(01)}  \\
   &=&  \tr{E_1^{{\rm L}}  
         \left( \CS^{(a)}(01)  \CS^{(b)}(01) + \CA^{(a)}(01)  \CA^{(b)}(01) \right) }. 
\end{eqnarray*}
We must calculate many traces in subsystems, for which 
the following formulas can be used ($p=a,b$):
\begin{eqnarray*}
 & & \trm \CM_3^{(p)}\CS^{(p)}(02)\CS^{(p)}(01) = \trm \CM_3^{(p)}\CA^{(p)}(02)\CA^{(p)}(01)  \\
 & & = \frac{1}{2}\tr{\CM_3^{(p)}(A^{(p)})^2} = \frac{1}{8}\dim V_\CM^{(p)},   \\
 & & \trm \CM_3^{(p)}\CS^{(p)}(02)\CA^{(p)}(01) = \trm \CM_3^{(p)}\CA^{(p)}(02)\CS^{(p)}(01)  \\
 & & = \frac{1}{2}\tr{\CM_3^{(p)}(D^{(p)})^2} = \frac{3}{8}\dim V_\CM^{(p)}. 
\end{eqnarray*}
The result is given by 
\begin{eqnarray*}
 & & \tr{E_1^{{\rm L}}\CS(01)}      \\ 
 &=& \frac{3}{8} \Big( 
         \alpha_1 \dim V_\CS^{(a)} \dim V_\CM^{(b)} + \alpha_2 \dim V_\CA^{(a)} \dim V_\CM^{(b)} \\ 
 & &    +\alpha_3 \dim V_\CM^{(a)} \dim V_\CA^{(b)} + \alpha_4 \dim V_\CM^{(a)} \dim V_\CS^{(b)}
                 \Big)            \\
 & &   +\frac{3}{32}(\beta_1+\beta_2)\dim V_\CM^{(a)} \dim V_\CM^{(b)}.
\end{eqnarray*}
It is clear that we should take the largest possible value 2/3 for 
the coefficients $\alpha$'s in order to maximize $\tr{E_1^{{\rm L}}\CS(01)}$.  
For $\beta_1$ and $\beta_2$, note that $\tr{E_1^{{\rm L}}\CS(01)}$ contains $\beta$'s in the 
form of $\beta_1+\beta_2$, and we can use the following inequalities:
\begin{eqnarray}
   \beta_1+\beta_2= 2\beta \le \frac{5}{4}\beta + \sqrt{ \frac{9}{16}\beta^2+\delta^2} \le 1.
\end{eqnarray}
Evidently, $\beta_1+\beta_2$ takes the maximum value 1 only when $\beta_1=\beta_2=1/2$. 
The maximum value of $\tr{E_1^{{\rm L}}\CS(01)}$ with separable POVM is thus given by 
\begin{eqnarray}
 & & \tr{E_1^{{\rm L}}\CS(01)}      \nonumber    \\ 
 &=& \frac{1}{4} \Big( 
         \dim V_\CS^{(a)} \dim V_\CM^{(b)} + \dim V_\CA^{(a)} \dim V_\CM^{(b)}  \nonumber \\ 
 & &    +\dim V_\CM^{(a)} \dim V_\CA^{(b)} + \dim V_\CM^{(a)} \dim V_\CS^{(b)}
                 \Big)              \nonumber    \\
 & &   +\frac{3}{32} \dim V_\CM^{(a)} \dim V_\CM^{(b)}.   \label{separable_trace_ES}
\end{eqnarray}
On the other hand, we have $\tr{E_1\CS(01)}= \frac{1}{4}\dim V_\CM$ for the global POVM element $E_1$.
We thus conclude that $\tr{E_1^{{\rm L}}\CS(01)} < \tr{E_1\CS(01)}$, since we have 
\begin{eqnarray}
  \dim V_\CM &=&  \dim V_\CS^{(a)} \dim V_\CM^{(b)} + \dim V_\CA^{(a)} \dim V_\CM^{(b)} 
                                   \nonumber \\
             & & +\dim V_\CM^{(a)} \dim V_\CA^{(b)} + \dim V_\CM^{(a)} \dim V_\CS^{(b)}
                                   \nonumber \\
             & & +\frac{1}{2} \dim V_\CM^{(a)} \dim V_\CM^{(b)}, 
\end{eqnarray}
which can be readily verified by an explicit calculation. This relation can be also 
understood from the viewpoint of inner (Kronecker) products of two representations of 
the symmetric group of order 3: the product of two mixed symmetric representations 
contains the totally symmetric and antisymmetric representations in addition to the 
mixed symmetric representation, whereas the product of the totally (anti)symmetric 
representation and the mixed symmetric representation is the mixed symmetric representation.  

\subsection{LOCC protocol}
Thus, the optimal separable POVM element $E_1^{{\rm L}}$ is given by Eq.(\ref{possible_E1}) with 
$\alpha_i=2/3$ and $\beta_i=1/2$:
\begin{eqnarray}
  E_1^{{\rm L}} &=&  \frac{2}{3} \Big( 
           \CS_3^{(a)} \otimes \CM_3^{(b)}\CA^{(b)}(02)
          +\CA_3^{(a)} \otimes \CM_3^{(b)}\CS^{(b)}(02)   \nonumber \\
      & & +\CM_3^{(a)}\CS^{(a)}(02) \otimes \CA_3^{(b)}
          +\CM_3^{(a)}\CA^{(a)}(02) \otimes \CS_3^{(b)} \Big)
                               \nonumber \\
      & & +\frac{1}{2} \big(  \CM_3^{(a)}\CS^{(a)}(02) \otimes \CM_3^{(b)}\CA^{(b)}(02)
                               \nonumber \\
      & & +\CM_3^{(a)}\CA^{(a)}(02) \otimes \CM_3^{(b)}\CS^{(b)}(02) \Big).
\end{eqnarray}
And the remaining elements are given as $E_2^{{\rm L}} = T(12)E_1^{{\rm L}}T(12)$ 
by the exchange symmetry, and $E_0^{{\rm L}} = 1-E_1^{{\rm L}}-E_2^{{\rm L}}$ by the 
completeness of POVM.
We can now show that this separable POVM $E_\mu^{{\rm L}}$ can be implemented by an LOCC protocol, 
which is summarized as follows:
\begin{itemize}
\item First, Alice and Bob determine the permutation symmetry of their local system: totally symmetric, 
mixed symmetric, or totally antisymmetric. This is done by the projective measurement with 
the set of orthogonal projectors $\{\CS_3^{(p)}, \CM_3^{(p)}, \CA_3^{(p)}\}$ of each party $p(=a,b)$. 
Note that the case $\CS_3^{(a)} \otimes \CA_3^{(b)}$ or $\CA_3^{(a)} \otimes \CS_3^{(b)}$ 
do not occur, since this would imply the total system is totally antisymmetric.
\item If their outcome is $\CS_3^{(a)} \otimes \CS_3^{(b)}$ or $\CA_3^{(a)} \otimes \CA_3^{(b)}$, 
Alice and Bob declare the inconclusive result, i.e., 0.
\item If one of the two parties $p(=a\ {\rm or }\ b)$ finds his or her local system is 
totally symmetric, $\CS_3^{(p)}$, and the system of the other party $q(\ne p)$ 
is found mixed symmetric, $\CM_3^{(q)}$, then party $q$ performs a POVM measurement: 
\begin{eqnarray*}
  e_1 & \equiv & \frac{2}{3}\CM_3^{(q)}\CA^{(q)}(02),     \\ 
  e_2 & \equiv & \frac{2}{3}\CM_3^{(q)}\CA^{(q)}(01),     \\ 
  e_0 & \equiv & \frac{1}{3}\CM_3^{(q)}\left(1+2A^{(q)}\right).
\end{eqnarray*}
Note that the set $\{e_1,e_2,e_0\}$ is a POVM since $e_\mu \ge 0,\ (\mu=0,1,2)$ and 
$\sum_\mu e_\mu = \CM_3^{(q)}$. 
The final identification result by Alice and Bob is the measurement outcome 
$\mu(=0,1,2)$ of party $q$.  
\item If one of the two parties $p(=a\ {\rm or }\ b)$ finds his or her local system is 
totally antisymmetric, $\CA_3^{(p)}$, and the system of the other party $q(\ne p)$ 
is found mixed symmetric, $\CM_3^{(q)}$, then party $q$ performs a POVM measurement: 
\begin{eqnarray*}
  e'_1 & \equiv & \frac{2}{3}\CM_3^{(q)}\CS^{(q)}(02),   \\ 
  e'_2 & \equiv & \frac{2}{3}\CM_3^{(q)}\CS^{(q)}(01),   \\ 
  e'_0 & \equiv & \frac{1}{3}\CM_3^{(q)}\left(1-2A^{(q)}\right).
\end{eqnarray*}
It is easily verified that the set $\{e'_1,e'_2,e'_0\}$ is also a POVM in $V_\CM^{(q)}$, and the 
final identification result of Alice and Bob is chosen to be the measurement outcome 
$\mu(=0,1,2)$ of party $q$. 
\item Finally, when the total system is found to be in $V_\CM^{(a)} \otimes V_\CM^{(b)}$, one of the two 
parties, say Alice, performs the following POVM measurement:
\begin{eqnarray*}
  && e_{11} \equiv \frac{1}{2}\CM_3^{(a)}\CA^{(a)}(02), \  
     e_{12} \equiv \frac{1}{2}\CM_3^{(a)}\CS^{(a)}(02), \\ 
  && e_{21} \equiv \frac{1}{2}\CM_3^{(a)}\CA^{(a)}(01), \
     e_{22} \equiv \frac{1}{2}\CM_3^{(a)}\CS^{(a)}(01).
\end{eqnarray*}
It is evident that the above set $\{e_{a_1a_2}\}_{a_1,a_2=1,2}$ forms a POVM in $V_\CM^{(a)}$. 
If Alice's outcome $a_1$ is equal to 1, Bob performs the projective measurement by 
the set of orthogonal projectors $\{f_b\}_{b=1,2}$:
\begin{eqnarray*}
   f_1 \equiv \CM_3^{(b)}\CS^{(b)}(02),\ 
   f_2 \equiv \CM_3^{(b)}\CA^{(b)}(02),
\end{eqnarray*}
otherwise, by the set of orthogonal projectors $\{f'_b\}_{b=1,2}$:
\begin{eqnarray*}
   f'_1 \equiv \CM_3^{(b)}\CS^{(b)}(01), \
   f'_2 \equiv \CM_3^{(b)}\CA^{(b)}(01).
\end{eqnarray*}
The final identification result $\mu$ of Alice and Bob is Alice's result $a_1$,  
if Alice's outcome $a_2$ coincides with Bob's outcome $b$. 
Otherwise, the final result is the inconclusive one, i.e., 0.
\end{itemize} 

Substituting explicit dimensions in Eq.(\ref{separable_trace_ES}), we obtain the optimal success 
probability with the LOCC protocol:   
\begin{eqnarray}
  p_{\max}^{{\rm L}} = \frac{1}{36d_ad_b(d_ad_b+1)} 
       \left( 11d_a^2d_b^2+d_a^2+d_b^2-13 \right ),
\end{eqnarray}
whereas the globally attainable success probability of Eq.(\ref{global_p}) in terms of dimensions 
$d_a$ and $d_b$ is given by  
\begin{eqnarray}
  p_{\max} = \frac{1}{3d_ad_b} (d_ad_b-1).
\end{eqnarray}
Although there is a finite gap between $p_{\max}$ and $p^{{L}}_{\max}$ as shown before, 
the numerical difference is not very large. 
For example, in the case of two-qubit bipartite system ($d_a=d_b=2$), 
the optimal LOCC protocol gives $p_{\max}^{{\rm L}}=19/80$, whereas 
the globally attainable probability is given by $p_{\max}=1/4$. The difference is only 1/80. 
In the limit of $d_a=d_b$ going to infinity, we find that $p_{\max}^{{\rm L}}$ approaches 11/36 and 
$p_{\max}$ approaches 1/3 with the difference 1/36.

\section{Concluding Remarks}
We have demonstrated that any LOCC scheme of the unambiguous (conclusive) identification of 
two bipartite pure states cannot attain the maximum success probability achieved by the 
global measurement. This contrasts remarkably with some known results for pure-state 
distinguishment problems with different settings. 
When classical knowledge of the states is given, it has been known that two bipartite pure 
states can be optimally discriminated, inconclusively \cite{Virmani01} and unambiguously 
(conclusively) \cite{Chen01,Chen02,Ji05}, by means of LOCC. 
It has also been shown \cite{Ishida08} that two bipartite pure states can be optimally 
identified by LOCC without classical knowledge, if one is allowed to make mistakes (inconclusive identification).  These results may be interpreted that there is no nonlocality 
in the distinguishment problems of two pure states. 
This paper provides an example of nonlocality in distinguishing two pure states. 
  
In this paper, we assumed the number of copies of each reference state is one.  
If unlimited number of copies of the reference states are available, one can always acquire 
complete classical information on the states, and the problem reduces to the standard unambiguous 
discrimination, where the LOCC scheme is known to perform as well as the global measurement scheme. 
It is an interesting problem to study the unambiguous identification of bipartite pure states 
when the number of copies of the reference states is finite but greater than one.

\end{document}